\documentclass[preprint,amsmath,showpacs]{revtex4}
\usepackage{graphicx}
\newcommand{\Deg}{{}^{\,\mbox{\scriptsize o}}}          % -  degree
\begin{document}
\draft
\title{\Large Three neutrino flavor oscillations and the atmospheric tau neutrinos}
\author{H. Athar\footnote{E-mail: athar@phys.cts.nthu.edu.tw}}
\address{Physics Division, National Center for Theoretical Sciences,
Hsinchu 300, Taiwan}
\date{\today}
\begin{abstract}
 Downward going atmospheric tau neutrino flux is estimated in the presence of
 three neutrino flavor oscillations  for 1 ${\rm GeV} \leq E\leq 10^{3}$ GeV. 
 The relative differences between  the three and purely two neutrino flavor
 oscillations  
 are elaborated. As an implication, the downward going
 atmospheric tau neutrino flux is compared with the galactic
 plane tau neutrino flux that is also estimated in the presence of
 three neutrino flavor  
 oscillations. It is pointed out that the galactic plane tau neutrino flux
 dominates over the downward going atmospheric tau neutrino 
 flux until $E \sim $ 10 GeV.  
\end{abstract}
\maketitle
\section{Introduction}
In order to perform a meaningful search of extra-atmospheric  neutrinos, the 
 atmospheric  neutrino background is needed to be known under
 various possible approximations \cite{Athar:2002rr}. 
 The search for astrophysical tau neutrinos is of special 
interest since  a positive search shall not only yield
the useful information from the surrounding cosmos 
but shall also corroborate the neutrino flavor mixing hypothesis \cite{Athar:2000yw}. 

In view of the presently available restricted ranges of the various neutrino 
mixing parameters  in the three neutrino flavor 
approximation \cite{Bahcall:2004ut}, as well as the recent relevant detector 
developments \cite{McDonald:2003xn}, it is of some interest to estimate 
 the resulting new atmospheric tau flux. 
 Such studies may also provide  some examples of 
 {\tt new windows} to study the cosmos around us, 
 that shall open up under the hypothesis of the neutrino
 oscillations.

Previously, there is {\tt no} attempt to estimate the atmospheric 
 tau neutrino flux
in the context of three neutrino flavor oscillations framework. 
 So far, the atmospheric tau neutrino flux is  estimated
in two neutrino flavor approximation only \cite{Athar:2004pb}. 
Recently, it is pointed out that the tau neutrino astronomy
 in the multi-GeV energy range
is possible, at least in principle.
 It is so, because the atmospheric $\nu_{\tau}$ flux is generally
suppressed compared to the atmospheric $\nu_{\mu}$ flux, as a result the
prospective observation of astronomical $\nu_{\tau}$ suffers 
 {\tt much less}
background than in the $\nu_{\mu}$ case.  The zenith angle ($\xi $) 
 dependence of the atmospheric background tau neutrino flux is also crucial
in determining the possible future prospects \cite{Athar:2004uk}. 
This is an example of the availability of a new 
 cosmic horizon because of the neutrino 
oscillations, and is true for {\tt tau neutrinos only} among 
 the three neutrino flavors. 

It is also pointed out that for 
 $0\Deg \leq \xi \leq 60\Deg $, the downward going
atmospheric tau neutrino flux is minimum, thus providing an opportunity
to search for extra-atmospheric tau neutrinos in this 
 zenith angle range, in the presence of neutrino 
 oscillations \cite{Athar:2004um}. The neutrino oscillation
effects are minimal for the down ward going atmospheric neutrinos. 
As a result, the oscillated atmospheric tau neutrino
flux provides a minimal background for prospective astronomical
 tau neutrino flux searches.

 Collectively, in these papers it is pointed out that 
{\tt contrary} to general expectations, the atmospheric neutrino
flux does not smear out the astronomical neutrino flux
on the Earth at low energy ($E < 10^{3}$ GeV) once the flavor composition of 
the incoming neutrino flux is taken into account.

In this paper, we estimate the atmospheric tau neutrino flux
in three neutrino flavor mixing approximation. This is an extension of 
 the above mentioned previous works in this direction. 
 As an implication of this
estimate, we compare it with galactic-plane tau neutrino flux
in the presence of neutrino oscillations. It is pointed out that 
the {\tt dominance} of galactic plane tau neutrino flux over the 
downward atmospheric tau neutrino flux still persists in 
 the multi-GeV energy range.

The considered energy range (1 ${\rm GeV} \leq E\leq 10^{3}$ GeV)
 is interestingly within the reach 
of presently operating detectors as well as those under active 
 planning \cite{McDonald:2003xn}. Tau neutrino flavor discrimination 
is an interesting challenge for the existing/forthcoming astrophysical 
neutrino telescopes.

This paper is organized as follows. In section II, the electron, muon and 
tau neutrino flux
originating from the Earth atmosphere is briefly discussed.
In section III, the neutrino oscillation effects
are studied for these. In section IV, 
an implication of our estimate for the atmospheric tau neutrinos is mentioned,
whereas in section V, conclusions are presented.
\section{Atmospheric neutrino flux}
Atmospheric neutrino flux arises when the incoming cosmic rays interact with the 
 air nuclei $A$, in the Earth atmosphere \cite{1962}.
For  1  GeV $\leq E \leq 10^{3}$ GeV, the $\pi $, the $K$ productions and
 their direct and indirect decays
 are the main sources of electron and muon neutrinos, 
 both being in the region of conventional
 neutrino production. The absolute normalization
of the conventional atmospheric neutrino flux is presently known to be no better
than (20$-$25)\% in the above energy range \cite{Battistoni:1999at}.
For present estimates, the non tau atmospheric neutrino flux is taken from
 Ref.~\cite{Honda:1995hz}. These are atmospheric 
 neutrino flux calculations in one dimension without geomagnetic field
 effects. At higher energy, the prompt muon 
neutrino production from $D$'s dominates over 
 the conventional one \cite{Volkova:gh}.

The atmospheric tau neutrino
flux arises mainly from production and the decay of $D_{S}$  and is
calculated in Ref.~\cite{Pasquali:1998xf,Athar:2001jw}.
 The Quark Gluon String Model (QGSM) is used in Ref.~\cite{Athar:2001jw} to
 model the $pA$ interactions.  The low energy
atmospheric tau neutrino flux is essentially isotropic \cite{Pasquali:1998xf}.
 For $E\leq  10^{2}$ GeV, the atmospheric tau neutrino flux is obtained by
 following the procedure given in Ref.~\cite{Pasquali:1998xf,Athar:2001jw}
 and re-scaling w.r.t new cosmic ray flux
spectrum, taking  it to be predominantly
the protons \cite{Honda:2004yz}.

\begin{figure}
\includegraphics[width=4.in]{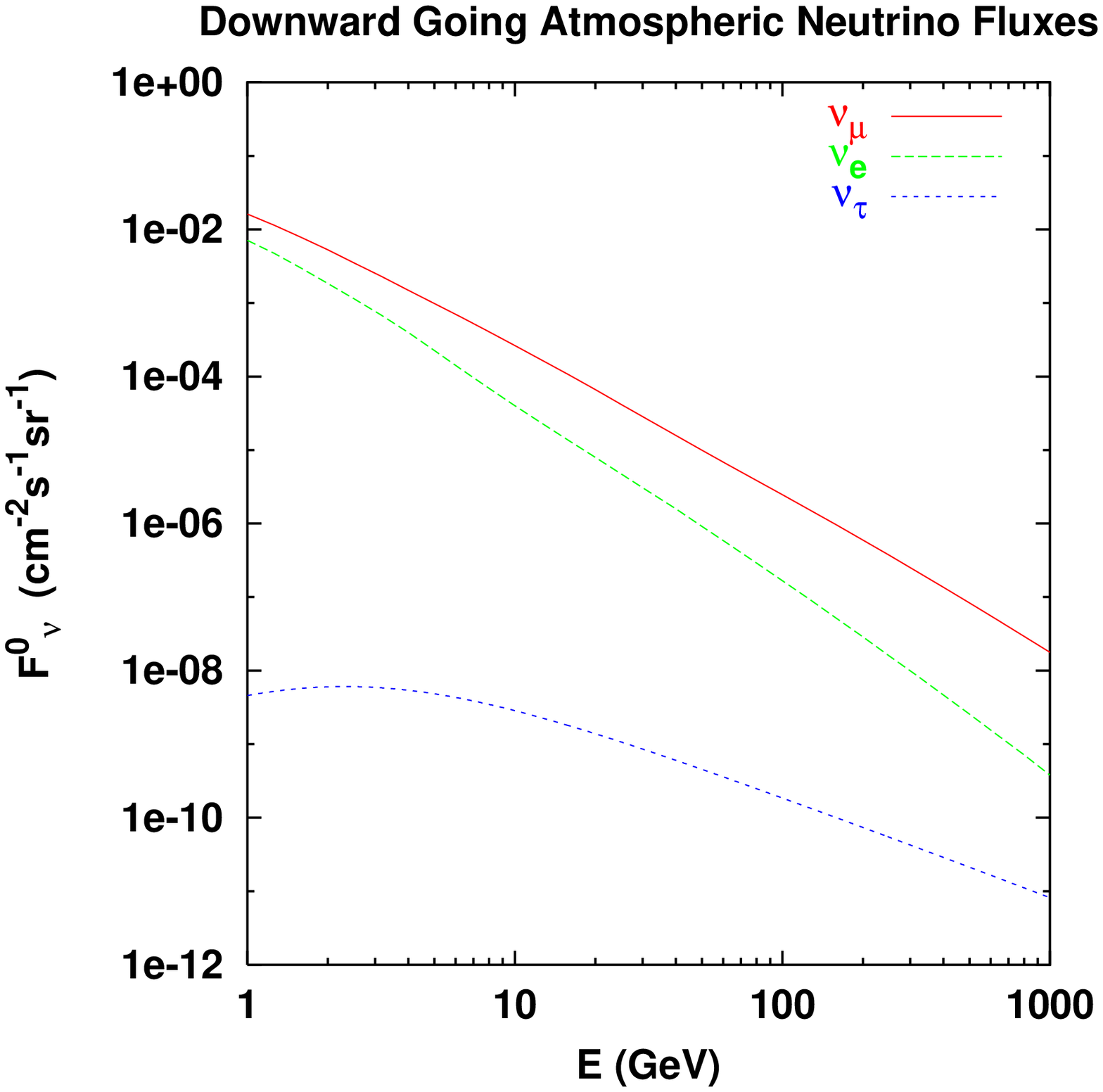}
\caption{The down ward going atmospheric electron, 
         muon and tau neutrino fluxes as a function of the neutrino energy. 
         {\em Absence} of neutrino oscillations is assumed here.
         More details are given in the text.}
\label{fig1}
\end{figure}

 For illustrative purpose, the three downward going ($\cos\xi =1$) 
 atmospheric neutrino fluxes 
 $F^{0}_{\nu}(E)\equiv {\rm d}^{2}N^{0}_{\nu}/{\rm d(log_{10}}E){\rm d}\xi$,
 in units of cm$^{-2}$s$^{-1}$sr$^{-1}$, 
estimated using the above description, are shown in Fig. \ref{fig1} as a
function of the neutrino energy. 
\section{Neutrino oscillation analysis: The Atmospheric tau neutrino flux}
We shall perform here the three neutrino flavor oscillation
analysis.
In the context of mixing in three flavors,  the
neutrino mixing parameters are : $\delta m^{2}_{12}$, $\delta m^{2}_{23}$,  
$\delta m^{2}_{13}$, $\theta_{12}$, $\theta_{23}$, $\theta_{13}$ and
the CP violating phase $\delta $. 
The $\delta m^{2}_{ij}\equiv |m^{2}_{i}-m^{2}_{j}|$, where $i,j=1,2,3$, is the 
 absolute difference of the mass squared value of the mass eigenstates and 
the three $\theta_{ij}$'s are the mixing angles. 

Presently, the $\delta m^{2}_{12}$ and $\theta_{12}$ are mainly determined  from 
the data analysis of the solar neutrino flux measurements \cite{Davis:2003kh}. The 
 neutrino mixing parameters inferred from solar neutrino flux studies 
 are recently confirmed by a terrestrial reactor experiment, namely Kamiokande 
 Liquid-scintillation Anti-Neutrino Detector (KamLAND) \cite{Araki:2004mb}.
The $\delta m^{2}_{23}$ and $\theta_{23}$ are mainly determined 
from the study of the atmospheric neutrinos \cite{Kajita:2000mr}. 
 The above neutrino mixing parameters inferred from atmospheric neutrino flux studies 
 are also recently confirmed by a terrestrial accelerator experiment, namely KEK to
 Kamiokande long-baseline neutrino experiment (K2K) \cite{Ishii:2004wu}.
 The $\theta_{13}$ is mainly constrained by the CHOOZ experiment \cite{Apollonio:1999ae}. 
 For some recent feasibility studies to determine more precisely the   
 $\theta_{13}$  and/or $\delta $, see \cite{Gandhi:2004md}.
The total range of $\delta m^{2}$ irrespective of neutrino flavor
in the context of three neutrino flavors is  
 $10^{-5}\, \, \, {\rm eV}^{2}\leq \delta m^{2} \leq 10^{-3}$ eV$^{2}$.

 Under the assumption that  $\delta =0 $, the 3$\times$3 
 Maki Nakagawa Sakata (MNS) mixing 
matrix U in standard parameterization connecting 
 the neutrino mass and flavor eigenstates reads \cite{Maki:1962mu}:
\begin{equation}
 U=\left( \begin{array}{ccc}
          c_{12}c_{13} & s_{12}c_{13} & s_{13}\\
          -s_{12}c_{23}-c_{12}s_{23}s_{13} & 
          c_{12}c_{23}-s_{12}s_{23}s_{13} & 
          s_{23}c_{13}\\
          s_{12}s_{23}-c_{12}c_{23}s_{13} & 
          -c_{12}s_{23}-s_{12}c_{23}s_{13} & 
          c_{23}c_{13}
          \end{array}
   \right),
\label{MNS}
\end{equation}
where $c_{ij}\equiv \cos\theta_{ij}$ and 
$s_{ij}\equiv \sin\theta_{ij}$. The presently available
empirical constraints for the various neutrino mixing parameters give  
 the elements of the above $U$ matrix as
\begin{eqnarray}
U&\equiv&\left(
\begin{array}{ccc}
U_{e1} & U_{e2} &  U_{e3}\\
U_{\mu 1} & U_{\mu 2} & U_{\mu 3} \\
U_{\tau 1} & U_{\tau 2} & U_{\tau 3}
\end{array}\right)
=\left( \begin{array}{ccc}
           0.84  &  0.54  & 0.10\\
          -0.44  &  0.56  & 0.71\\
           0.32  & -0.63  & 0.71
          \end{array}
   \right).
\label{U}
\end{eqnarray}
A recent short summary  of the model predictions for the
 expected magnitude of $|U_{e3}|$ 
  is given in Ref. \cite{Joshipura:2004ws}. 
 Using Eq. (\ref{MNS}),  
 the neutrino flavor oscillation probability formula 
is \cite{Athar:2002uj}
\begin{equation}
 P(\nu_{\alpha} \to \nu_{\beta} ;E, \xi)
 \equiv P_{\alpha \beta}=\sum^{3}_{i=1}
 U_{\alpha i}^{2}U_{\beta i}^{2}+
 \sum_{i\neq j}U_{\alpha i}U_{\beta i}U_{\alpha j}U_{\beta j}
 \cos \left(\frac{2L}{L_{ij}}\right),
\label{prob}
\end{equation}
where $\alpha, \beta =e, \mu, \tau$ and $L_{ij}\simeq 4E/\delta m^{2}_{ij}$ 
 is the neutrino oscillation length. 
 The $L$  in Eq. (\ref{prob}) is the neutrino flight length. 
In the Earth atmosphere, it  can be estimated using the relation
\begin{equation}
 L=\sqrt{(h^{2}+2R_{\oplus}h)+(R_{\oplus}\cos \xi)^{2}}-R_{\oplus} \cos \xi,
\label{len}
\end{equation}
where $\xi $ is the zenith angle. 
The $L$ is essentially the distance between the detector and the height
 at which the atmospheric neutrinos are produced.
 The $R_{\oplus}\simeq 6.4 \cdot 10^{3}$ km is the Earth radius, and  
 for instance, $h_{\mu}=15$ km
 is the mean altitude at which the atmospheric muon neutrinos are produced.
 In general, $h$ is not only a function of the zenith angle $\xi$, the neutrino flavor
 but also the neutrino energy \cite{Gaisser:1997eu}. 

\begin{figure}
\includegraphics[width=4.in]{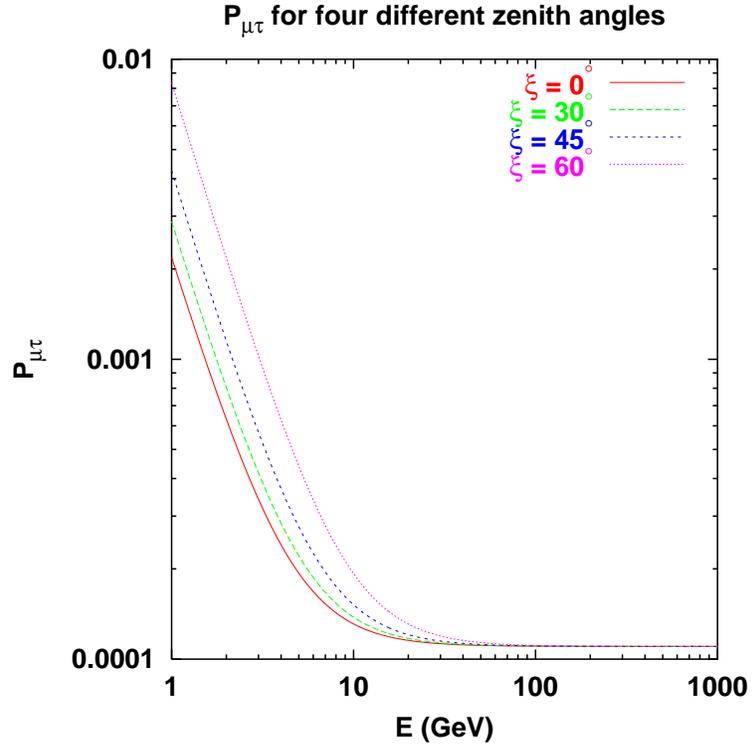}
\caption{An illustration of $P_{\mu \tau}$ for four different $\xi$ values,
  as a function of the neutrino energy. Three neutrino flavor mixing is 
 assumed here.}
\label{fig2}
\end{figure}

We consider the zenith angle range between $0^{\Deg}$ and $60^{\Deg}$.
Because in this energy range i) as pointed out in Ref. \cite{Athar:2004um}, 
 the atmospheric
neutrino oscillation effect is minimal, ii) the Earth curvature effects
 can be ignored, and iii) more relevant for present discussion is that in this
zenith angle range, the matter effects in neutrino oscillations are negligible  as
  the Earth atmosphere matter density traversed by neutrino satisfies: 
 $\rho < \rho_{\rm res}$ where 
$\rho_{\rm res}\equiv  m_{N}\Delta \cos 2\theta/\sqrt{2}G_{F}$, 
 with $\Delta \equiv \delta m^{2}/2E$ ranges between $10^{-17}$ eV and
 $10^{-12}$ eV,
 for the entire considered $\delta m^{2}$ and $E$ ranges.
Fig. \ref{fig2} shows an 
 example of $P_{\mu \tau}$ using Eq. (\ref{prob}) 
 for four different $\xi$ values where $\xi =0^{\Deg}$ corresponds to the downward going
atmospheric neutrinos. A relevant remark useful for  
 what follows next is that for $E < 20$ GeV, the $P_{\mu \tau}$ is larger
by a factor of up to $\sim 8$ for larger $\xi $, owing to the fact that for
larger $\xi $, neutrinos traverse larger distance in Earth atmosphere.

\begin{figure}
\includegraphics[width=4.in]{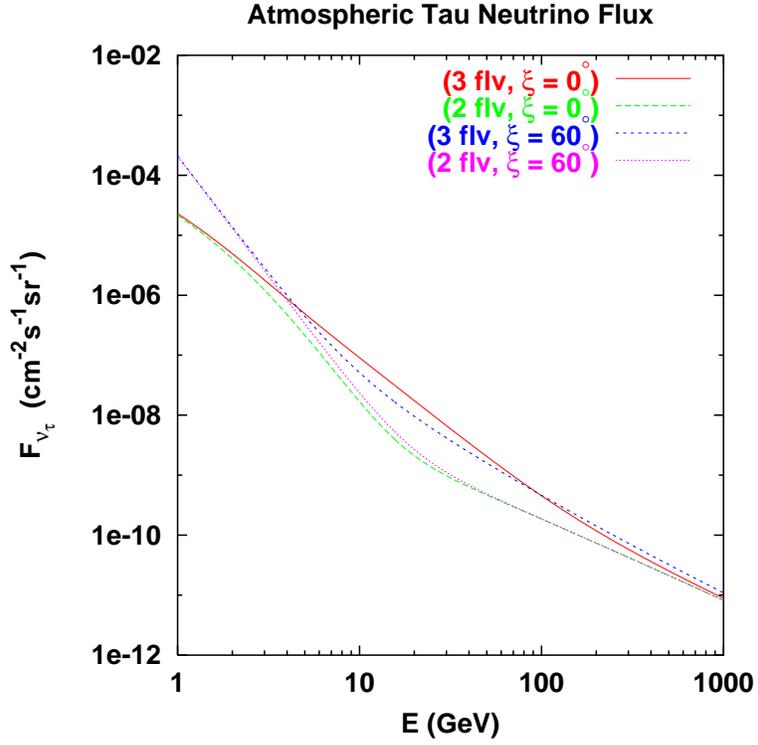}
\caption{The downward going atmospheric tau 
 neutrino flux in the three and the two
 neutrino mixings for $\xi=0^{\Deg}$ and $\xi=60^{\Deg}$ 
 as a function of the neutrino energy.}
\label{fig3}
\end{figure}

In the limit 
$\theta_{12},\theta_{13}\to 0$, we obtain the commonly used two neutrino
flavor oscillation probability formula for 
 $P(\nu_{\mu} \to \nu_{\tau} ;E, \xi)$ as
\begin{equation}
 P_{\mu \tau}=\sin^{2}2\theta_{23}\cdot
 \sin^{2}\left(1.27 
 \frac{\delta m^{2}_{23}({\rm eV}^{2})L({\rm km})}{E({\rm GeV})}\right),
\label{2flv}
\end{equation}
where $\sin2\theta_{23}=2c_{23}s_{23}$. 
 The neutrino flux $F_{\nu_{\alpha}}(E, \xi)$, arriving at the detector,
 in the presence of neutrino 
 oscillations is estimated using 
\begin{equation}
 F_{\nu_\alpha}(E,\xi)=\sum_{\beta} P_{\alpha \beta}(E, \xi)
 F^{0}_{\nu_\beta}(E, \xi),
\label{tot}
\end{equation}
where $F^{0}_{\nu_\beta}(E, \xi)$ are taken according to discussion in section II. 
The $P_{\alpha \beta}(E, \xi)$ is a 3$\times$3 matrix obtainable
 using Eq. (\ref{prob}). The unitarity conditions such as 
$1-P_{ee}(E,\xi)=P_{e\mu}(E, \xi)+P_{e\tau}(E, \xi)$ are 
implemented at each $E$ and $\xi $ at which these are evaluated.
Fig. \ref{fig3} shows  comparison of the atmospheric tau neutrino
flux for $\xi = 0^{\Deg}$ and $\xi = 60^{\Deg}$ in three and 
 two neutrino flavor
mixings as a function of the neutrino energy. Note the difference in
the slope in three and two neutrino flavor approximations
because of the presence of second $\delta m^{2}$ scale in
three flavor mixing relative to only one $\delta m^{2}$ scale 
 in purely two neutrino flavor approximation.
 The atmospheric tau neutrino flux for larger zenith angles
is larger relative to smaller zenith angles for $E < 20$ GeV
 (see Fig. \ref{fig2} also).

The downward going atmospheric tau neutrino flux 
in the presence of three neutrino flavor oscillations can be
parameterized for 1 GeV $\leq E \leq 10^{2}$ GeV  as 
\begin{equation}
 F_{\nu_{\tau}}(E) = A-Bx^{-1}
 +Cx^{-2}+Dx^{-3}-Ex^{-4},
\end{equation}
where the coefficients are $A=4.20865\cdot 10^{-11}$, $B=1.57226\cdot 10^{-8}$,
 $C=5.37912\cdot 10^{-6}$, $D=3.97309\cdot 10^{-5}$ and $E=2.21708\cdot 10^{-5}$  
 with $x=E/{\rm GeV}$. The $F_{\nu_{\tau}}(E)$ is in units of
cm$^{-2}$s$^{-1}$sr$^{-1}$ . 
\section{An implication: Comparison with the Galactic Plane Tau
         Neutrino Flux}
In this section, we shall estimate the galactic plane
tau neutrino flux with three neutrino flavor mixing.
This is to {\em illustrate} a {\tt new} opportunity we may
have to study cosmos having estimated the related atmospheric
tau neutrino background.

The galactic plane electron and muon neutrino flux
 is calculated in Ref.~\cite{Stecker:1978ah}, whereas
the galactic plane tau neutrino flux is calculated in Ref.~\cite{Athar:2001jw}.
 These calculations consider $pp$ interactions inside the galaxy with
 target proton number density $\sim $ 1/cm$^{3}$ along the galactic plane,
 under the assumption that the cosmic ray flux spectrum  
 in the galaxy is constant
 at its locally observed  value.
 We emphasize that because of above uncertainties, the
galactic plane neutrino fluxes should only be considered as
reference upper limits.

Following Ref.~\cite{Athar:2001jw}, the galactic plane
 neutrino flux for 1 GeV $\leq E\leq  10^{3}$
 GeV is obtained by re-scaling w.r.t new cosmic ray flux
spectrum. The tau neutrino production is rather suppressed in the
galactic plane relative to muon neutrino production.
 In general, the  non tau neutrino flux is larger than the tau
neutrino flux for $E\leq  10^{3}$ GeV from the above two sites.

In order to estimate the oscillation effect, 
 the distance $L$ for galactic plane neutrinos
 can be taken as $\sim $ 5 kpc, where 1 pc $\sim  3\times 10^{13}$ km.
 This imply that
$L_{ij} \ll L$, namely the galactic plane   neutrino flux oscillate
before reaching the earth. Thus, the galactic plane neutrino flux is averaged out
 for the whole range of $\delta m^{2}$ in the entire considered energy range.
 Under the 
assumption of averaging over rapid oscillations,
we obtain from Eq. (\ref{prob})
\begin{equation}
 P_{\alpha \beta}\simeq \sum^{3}_{i=1}
 U_{\alpha i}^{2}U_{\beta i}^{2}.
\label{prob2}
\end{equation}
The $3\times 3\, \, \, P$ matrix using  Eq. (\ref{U})
in above Eq. is
\begin{eqnarray}
P&\equiv&\left(
\begin{array}{ccc}
P_{ee} & P_{e\mu} &  P_{e\tau}\\
P_{e\mu} & P_{\mu \mu } & P_{\mu \tau} \\
P_{e \tau } & P_{\mu \tau } & P_{\tau \tau}
\end{array}\right)
=\left( \begin{array}{ccc}
          0.59  & 0.23  & 0.20\\
          0.23  & 0.39  & 0.39\\
          0.20  & 0.39  & 0.42
          \end{array}
   \right).
\label{gal-p}
\end{eqnarray}
Note that this $P$ matrix is independent of not only $\delta m^{2}$ but
also $E$ (and $\xi $). 
A rather similar $3\times 3\, \, \, P$ matrix was obtained 
 for both vanishing $\delta $ and $\theta_{13}$ in Ref. \cite{Athar:2000ak}. 

\begin{figure}[t]
\hglue -8.cm
\includegraphics[width=3in]{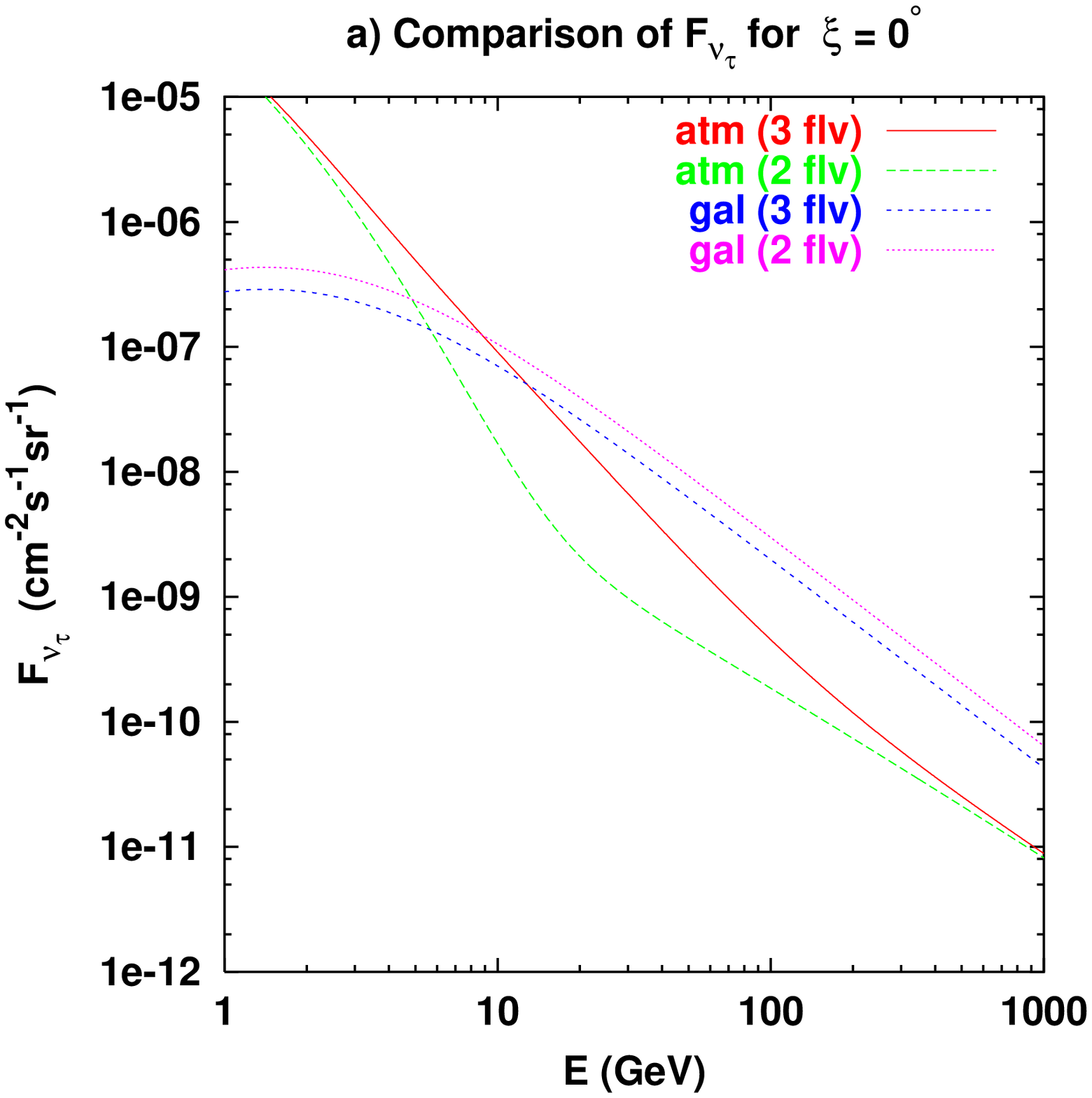}
\vglue -7.75cm 
\hglue 8.4cm \includegraphics[width=3in]{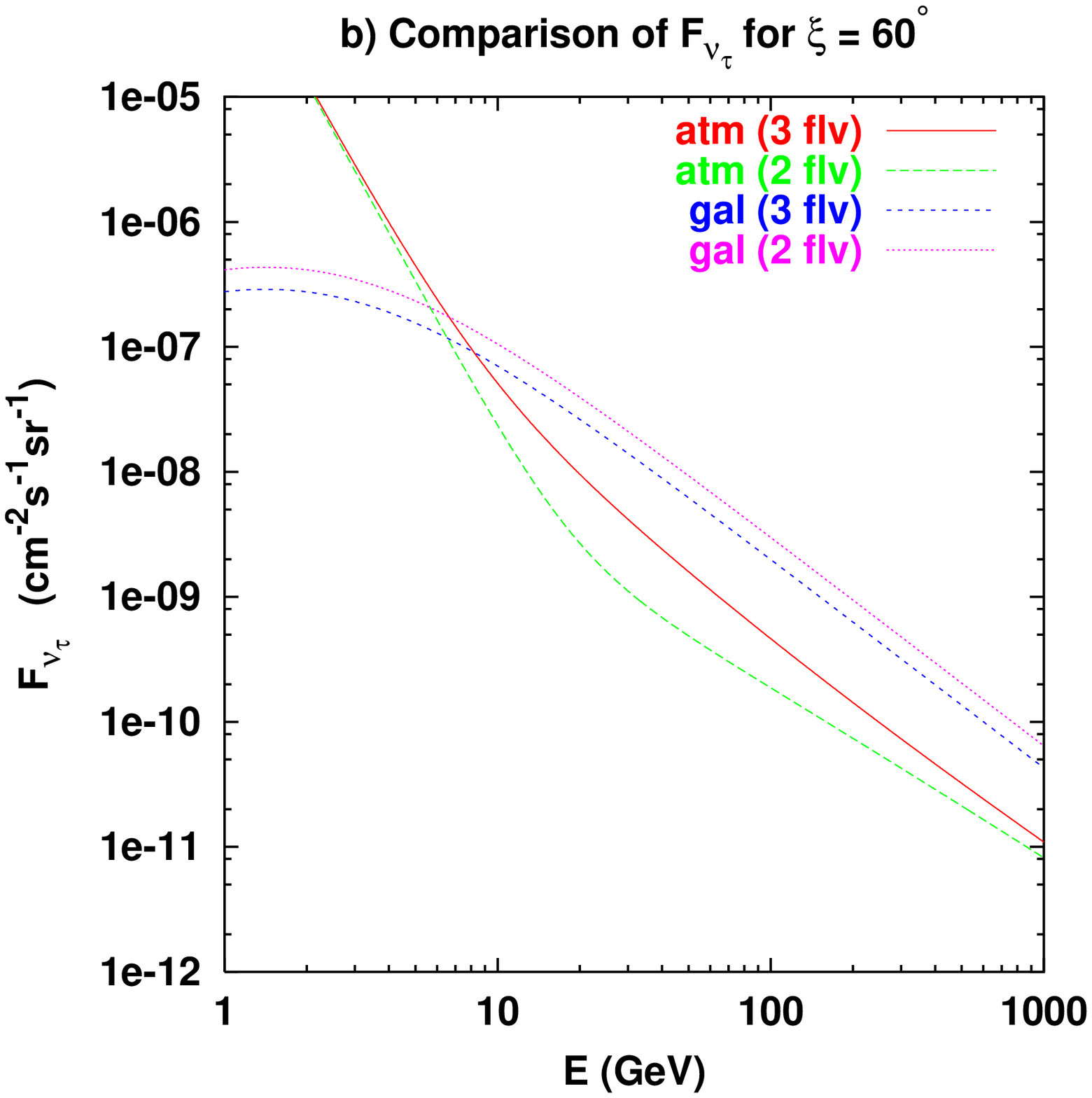}
\caption{Left panel: The comparison of the galactic plane and the
 downward going atmospheric tau neutrino flux for $\xi = 0^{\Deg}$ 
 as a function of the neutrino energy. Three
 flavor neutrino mixing is assumed. Also shown is two neutrino
 flavor mixing results with maximal 23 mixing.
 Right panel: Same as left panel but for $\xi = 60^{\Deg}$.}
\label{fig4}
\end{figure}

 Using Eq. (\ref{tot}) and Eq. (\ref{gal-p}), the galactic plane 
tau neutrino flux is estimated in the presence of three neutrino mixing. 
 It is then compared with the atmospheric tau neutrino flux 
 in Fig. \ref{fig4}. The comparison 
 includes the two neutrino flavor approximation effect with 
 maximal 23 mixing also. 
 From the figure, it can be seen
that the galactic plane/non-atmospheric tau neutrino flux starts {\tt dominating} over
the downward going atmospheric tau neutrino flux even for
$E$ as low as 10 GeV in the presence of neutrino oscillations, 
 depending upon the incident zenith angle for Atmospheric neutrinos
 (the galactic plane neutrino flux is independent of the zenith angle). 
 This is a very specific feature of {\tt tau neutrinos},
 and is absent for electron and muon neutrinos. This specific
behavior has to do with the {\tt neutrino oscillations}. 
The galactic plane tau neutrino flux in three neutrino mixing is smaller by a 
factor $\sim 2/3$ relative to two neutrino flavor mixing, irrespective of the neutrino 
 energy.  

Fig. \ref{fig4} indicates that zenith angle
dependence of the total tau neutrino flux can at least in
principle help to distinguish between atmospheric and
 non-atmospheric tau neutrino flux. The galactic plane
 tau neutrino flux dominance is
 {\tt independent} of number of oscillating neutrino flavors (two or three).
 The galactic tau
neutrino flux transverse to the galactic plane is
three to four orders of magnitude smaller than the
galactic plane one \cite{Athar:2001jw}.

\begin{figure}
\includegraphics[width=4.in]{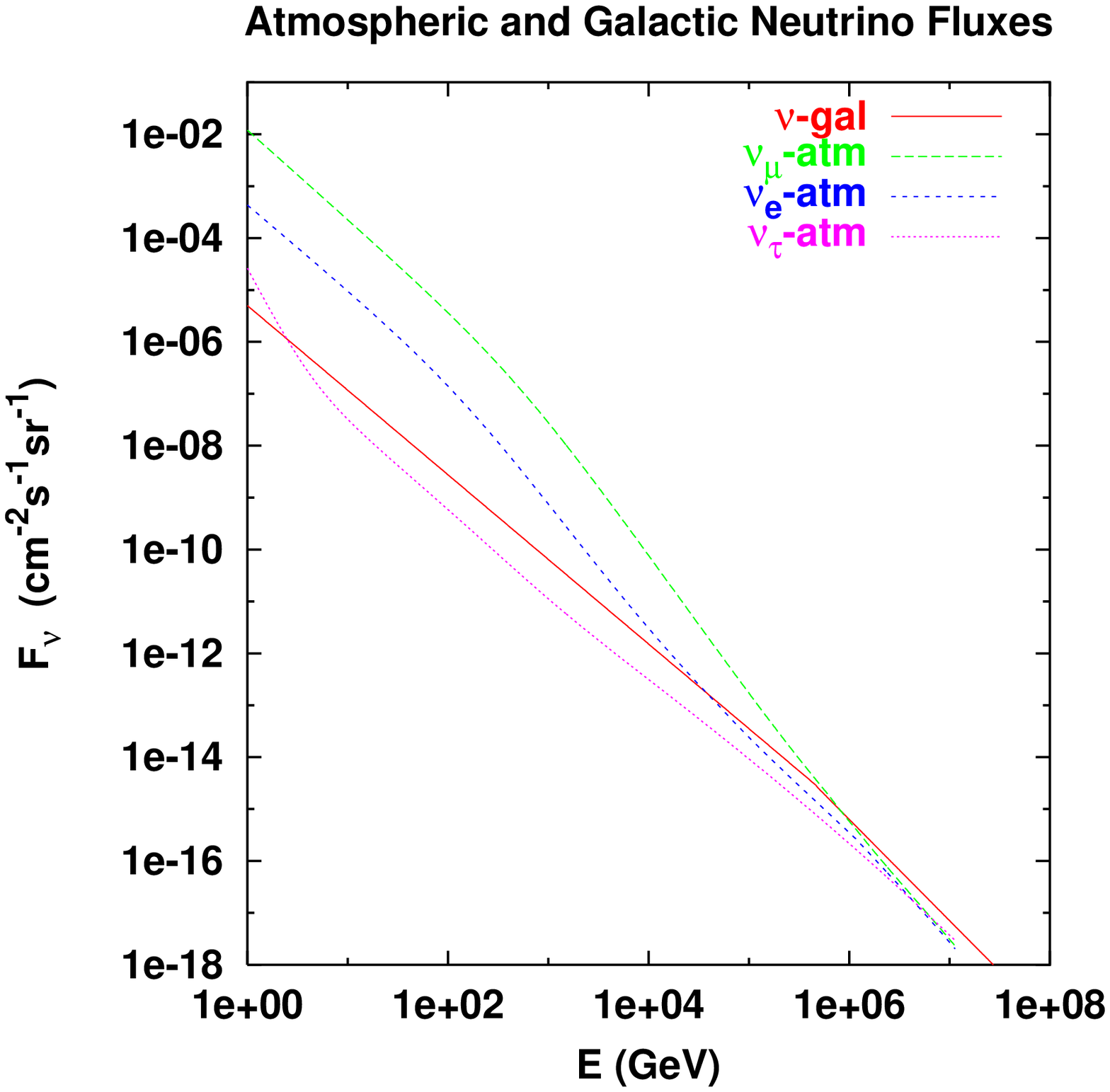}
\caption{The down ward going atmospheric electron, 
         muon and tau neutrino fluxes as a function of the neutrino energy
         in the presence of three neutrino flavor mixing. The 
         galactic plane neutrino flux estimated in three neutrino 
         neutrino flavor mixing is also shown for comparison.}
\label{fig5}
\end{figure}

Fig. \ref{fig5} shows the three downward going atmospheric
neutrino fluxes in the presence of three neutrino flavor mixing, 
 along with the corresponding galactic plane neutrino fluxes. 
Here the contribution from the charm production is
also taken into account \cite{Gondolo:1995fq}. The galactic plane
three neutrino flavor fluxes are approximately equal because of
Eq. (\ref{gal-p}). The unique behavior of atmospheric oscillated tau neutrino
flux relative to electron and muon neutrino flux is evident. 
The {\tt oscillated} galactic plane tau neutrino flux thus {\tt dominates} until 
 approximately 10 GeV over the corresponding atmospheric neutrino 
background. On the other hand, the same occurs at E $\sim 10^{5}$ GeV and
E $\sim 10^{6}$ GeV for electron and muon neutrino fluxes in the 
presence of three neutrino flavor mixing from the two astrophysical
sources under discussion. The change in the slope of the 
 atmospheric electron and muon neutrino 
fluxes is a reflection of the corresponding change in the slope of
cosmic-ray flux.

	In general, the three downward atmospheric neutrino fluxes
provide an {\tt energy dependent} background to the incoming 
astronomical neutrino flux in the {\tt presence} of neutrino oscillations. 
 This observation may have some relevance
for the forthcoming neutrino telescopes with the prospective neutrino flavor
discrimination capabilities.

The
galactic plane tau neutrino flux for 1 GeV $\leq E \leq 10^{3}$ GeV in the
presence of three neutrino flavor mixing can be parameterized as
\begin{equation}
 F_{\nu_{\tau}}(E) =
 4.38\cdot 10^{-6} \cdot E^{1.07}\left[ E+2.15\exp({-0.21\sqrt{E}}) \right]^{-2.74},
\label{parameterize}
\end{equation}
where $F_{\nu_{\tau}}(E)$ is in units of
cm$^{-2}$s$^{-1}$sr$^{-1}$ and on r.h.s. $E$ is in units of GeV.

We have estimated the galactic plane tau
neutrino induced shower event rate for the forthcoming  Mega ton (Mt) class
of detectors \cite{uno}, to indicate the limited prospects offered
by below TeV (1 TeV = $10^{3}$ GeV) tau neutrino astronomy to search for
extra-atmospheric astrophysical neutrino sources in this energy range, 
 as envisaged  presently.
 The tau neutrino event rate $N_{\rm event}$ for 
 10 GeV $\leq E \leq 10^{3}$ GeV can be 
 approximately estimated as
\begin{equation}
 N_{\rm event}=\int^{E^{\rm max}}_{E^{\rm min}}
 F_{\nu_{\tau}}(E)  \cdot
 \sigma_{\nu_{\tau}\to {\rm shower}}(E) \frac{{\rm d}E}{E}.
\label{rate}
\end{equation}
The $F_{\nu_{\tau}}(E) $ is the galactic plane tau neutrino flux in the
 presence of neutrino oscillations in units of 
cm$^{-2}$s$^{-1}$sr$^{-1}$ and is given by Eq. (\ref{parameterize}). 
The $\sigma_{\nu_{\tau}\to {\rm shower}}(E)$
is $\sigma^{\rm CC}_{\nu_{\tau}N}(E)/{\rm Mt}$, where one megaton of water contains
 $\sim 5.5\cdot 10^{35}$ nucleons \cite{book}. The $\sigma_{\nu_{\tau}\to {\rm shower}}(E)$
 is taken from the web site of Cern Neutrinos to Gran Sasso (CNGS) experiment \cite{cngs}.  
 The $N_{\rm event}$ is found to be in the range 1 to 10 in units of 
 ${\rm Mt}^{-1}{\rm yr}^{-1}$ over  
 $2\pi \cdot {\rm sr}$ with a  3 to 5 years of data taking. 
\section{Conclusions}
Atmospheric tau neutrino flux is estimated for 1 GeV $\leq E \leq 10^{3}$ GeV
 in the presence of neutrino oscillations. The three  neutrino flavor
oscillation study is carried out. The relative difference between 
two and three neutrino flavor oscillation analysis is elaborated.

As an implication of this study, the  atmospheric tau
neutrino flux is compared with the galactic plane tau neutrino 
flux with three neutrino mixing. 
 It is pointed out that the dominance of galactic plane
tau neutrino flux persists over the downward going ($\xi =0\Deg$) atmospheric tau
neutrino flux for $E\geq 10$ GeV relative to $E\geq 8$ GeV in two 
flavor approximation. The same  dominance shifts to $E\geq 10$ GeV
relative to $E\geq 9$ GeV  for
$\xi = 60\Deg$. 

In the  multi-GeV energy range, the dominance of the galactic 
 plane tau neutrino flux
over the atmospheric tau neutrino flux for incident zenith angle
between $0^{\Deg}$ and $60^{\Deg}$ in the presence of neutrino
oscillation both in two and three neutrino mixing is {\tt unique} to
tau neutrinos only. The galactic electron and muon neutrino
flux dominance starts only for $E\geq  10^{5}$ GeV irrespective
of mixed (two or three) neutrino flavors. 

Summarizing, a possibility of a {\tt neutrino flavor dependent new window} in
low energy neutrino astronomy seems to exist. Our present study is 
an attempt to justify this existence. Namely, for the incident 
neutrino energy range  between 10s of GeV up to 1000s of GeV, 
 in which the presently data taking large neutrino detectors 
 such as the Antarctic Muon And Neutrino Detector Array (AMANDA) 
 experiment \cite{amanda}
at the south pole and the Baikal experiment \cite{baikal} in the lake Baikal in Russia
are not (yet) optimized to perform extra-atmospheric neutrino astronomy, 
the upcoming Mega ton class of detectors (or even the currently existing ones) 
should at least in principle be able to do possibly
a meaningful neutrino astronomy based on prospective neutrino flavor
identification.  
\section*{Acknowledgement}
The author thanks Physics Division of NCTS for support.
\pagebreak
\pagebreak
\pagebreak
\pagebreak

\begin{thebibliography}{99}
%
%
%
\bibitem{Athar:2002rr}
For a general introduction to this topic, see,
 H. Athar, in {\it Proceedings of the IAU 8th Asian-Pacific Regional Meeting, 
Tokyo, Japan, 2002}, edited by S. Ikeuchi, J. Hearnshaw and T. Hanawa 
(Astronomical Society of the Pacific, San Francisco, 2003), p. 323
%``High energy astrophysical neutrinos,''
 [arXiv:hep-ph/0209130].
%%CITATION = HEP-PH 0209130;%%
 For a recent brief review article, see, {\it ibid}., 
%``Mixed high energy neutrinos from cosmos,''
 Chin.\ J.\ Phys.\  {\bf 42}, 1 (2004). 
%%CITATION = HEP-PH 0308188;%%
%
%
%
\bibitem{Athar:2000yw}
H.~Athar, M.~Je\.{z}abek and O.~Yasuda,
%``Effects of neutrino mixing on high-energy cosmic neutrino flux,''
Phys.\ Rev.\ D {\bf 62}, 103007 (2000). 
%%CITATION = HEP-PH 0005104;%%
For a recent discussion, see, 
 S.~Pakvasa,
%``Neutrino properties from high energy astrophysical neutrinos,''
Mod.\ Phys.\ Lett.\ A {\bf 19}, 1163 (2004).
%%CITATION = HEP-PH 0405179;%%
%
%
%
\bibitem{Bahcall:2004ut}
See, for instance, J.~N.~Bahcall, M.~C.~Gonzalez-Garcia and C.~Pena-Garay,
%``Solar neutrinos before and after Neutrino 2004,''
JHEP {\bf 0408}, 016 (2004).
%%CITATION = HEP-PH 0406294;%%
%
%
%
\bibitem{McDonald:2003xn}
A.~B.~McDonald {\it et al}.,
%``Astrophysical neutrino telescopes,''
Rev.\ Sci.\ Instrum.\  {\bf 75}, 293 (2004).
%%CITATION = ASTRO-PH 0311343;%%
%
%
%
\bibitem{Athar:2004pb}
H.~Athar,
%``Atmospheric and galactic tau neutrinos,''
Mod.\ Phys.\ Lett.\ A {\bf 19}, 1171 (2004).
%%CITATION = HEP-PH 0401242;%%
%
%
%
\bibitem{Athar:2004uk}
H.~Athar and C.~S.~Kim,
%``GeV to TeV astrophysical tau neutrinos,''
Phys.\ Lett.\ B {\bf 598}, 1 (2004).
%%CITATION = HEP-PH 0407182;%%
%
%
%
\bibitem{Athar:2004um}
H.~Athar, F.~F.~Lee and G.~L.~Lin,
%``Tau neutrino astronomy in GeV energies,''
arXiv:hep-ph/0407183.
%%CITATION = HEP-PH 0407183;%%
%
%
%
\bibitem{1962}
G. T. Zatsepin and V. A. Kuzmin,
 Sov. Phys. JETP {\bf 14}, 1294 (1962). 
%%CITATION = NONE;%%
 For a recent review article, see, 
T.~K.~Gaisser and M.~Honda,
%``Flux of atmospheric neutrinos,''
 Ann.\ Rev.\ Nucl.\ Part.\ Sci.\  {\bf 52}, 153 (2002).
%%CITATION = HEP-PH 0203272;%%
%
%
%
\bibitem{Battistoni:1999at}
 See, for instance,
 G.~Battistoni {\it et al}.,
%``A 3-dimensional calculation of atmospheric neutrino flux,''
 Astropart.\ Phys.\  {\bf 12}, 315 (2000).
%%CITATION = HEP-PH 9907408;%%
%
%
%
\bibitem{Honda:1995hz}
M.~Honda, T.~Kajita, K.~Kasahara and S.~Midorikawa,
%``Calculation of the flux of atmospheric neutrinos,''
 Phys.\ Rev.\  {\bf D52}, 4985 (1995);
%%CITATION = HEP-PH 9503439;%%
 V.~Agrawal, T.~K.~Gaisser, P.~Lipari and T.~Stanev,
%``Atmospheric neutrino flux above 1 GeV,''
  Phys.\ Rev.\    {\bf D53}, 1314 (1996).
%%CITATION = HEP-PH 9509423;%%
%
%
%
\bibitem{Volkova:gh}
H.~Inazawa and K.~Kobayakawa,
%``The Production Of Prompt Cosmic Ray Muons And Neutrinos,''
 Prog.\ Theor.\ Phys.\  {\bf 69}, 1195 (1983).
%%CITATION = PTPKA,69,1195;%%
 For a recent discussion, see, 
 H.~Athar, K.~Cheung, G.~L.~Lin and J.~J.~Tseng,
%``The high-energy galactic tau neutrino flux and its atmospheric background,''
Eur.\ Phys.\ J.\ C {\bf 33}, S959 (2004), and references cited therein.
%%CITATION = ASTRO-PH 0311586;%%
%
%
%
\bibitem{Pasquali:1998xf}
L.~Pasquali and M.~H.~Reno,
%``Tau neutrino fluxes from atmospheric charm,''
 Phys.\ Rev.\  {\bf D59}, 093003 (1999).
%%CITATION = HEP-PH 9811268;%%
%
%
%
\bibitem{Athar:2001jw}
H.~Athar, K.~Cheung, G.~L.~Lin and J.~J.~Tseng,
%``High-energy cosmic-ray tau neutrino flux from the Milky Way,''
 Astropart.\ Phys.\  {\bf 18}, 581 (2003).
%%CITATION = HEP-PH 0112222;%%
%
%
%
\bibitem{Honda:2004yz}
M.~Honda, T.~Kajita, K.~Kasahara and S.~Midorikawa,
%``A new calculation of the atmospheric neutrino flux in a 3-dimensional
%scheme,''
Phys.\ Rev.\ D {\bf 70}, 043008 (2004).
%%CITATION = ASTRO-PH 0404457;%%
%
%
%
\bibitem{Davis:2003kh}
R.~Davis,
%``Nobel Lecture: A half-century with solar neutrinos,''
Rev.\ Mod.\ Phys.\  {\bf 75}, 985 (2003).
%%CITATION = RMPHA,75,985;%%
%
%
%
\bibitem{Araki:2004mb}
T.~Araki {\it et al.}  [KamLAND Collaboration],
%``Measurement of neutrino oscillation with KamLAND: Evidence of spectral
%distortion,''
arXiv:hep-ex/0406035.
%%CITATION = HEP-EX 0406035;%%
%
%
%
\bibitem{Kajita:2000mr}
T.~Kajita and Y.~Totsuka,
%``Observation of atmospheric neutrinos,''
Rev.\ Mod.\ Phys.\  {\bf 73}, 85 (2001).
%%CITATION = RMPHA,73,85;%%
%
%
%
\bibitem{Ishii:2004wu}
T.~Ishii  [K2K Collaboration],
%``Recent K2K results,''
arXiv:hep-ex/0406055;
%%CITATION = HEP-EX 0406055;%%
 E.~Aliu [K2K Collaboration],
%``Evidence for muon neutrino oscillation in an accelerator-based experiment,''
arXiv:hep-ex/0411038.
%%CITATION = HEP-EX 0411038;%%
%
%
%
\bibitem{Apollonio:1999ae}
M.~Apollonio {\it et al.}  [CHOOZ Collaboration],
%``Limits on neutrino oscillations from the CHOOZ experiment,''
Phys.\ Lett.\ B {\bf 466}, 415 (1999).
%%CITATION = HEP-EX 9907037;%%
%
%
%
\bibitem{Gandhi:2004md}
R.~Gandhi {\it et al}.,
%``Large matter effects in nu/mu $\to$ nu/tau oscillations,''
arXiv:hep-ph/0408361;
%%CITATION = HEP-PH 0408361;%%
K.~B.~McConnel and M.~H.~Shaevitz,
%``Comparisons and Combinations of Reactor and Long-Baseline Neutrino
%Oscillation Measurements,''
arXiv:hep-ex/0409028;
%%CITATION = HEP-EX 0409028;%%
H.~Sugiyama, O.~Yasuda, F.~Suekane and G.~A.~Horton-Smith,
%``Systematic limits on sin^2{2theta_{13}} in neutrino oscillation experiments
%with multi-reactors,''
arXiv:hep-ph/0409109;
%%CITATION = HEP-PH 0409109;%%
 Q.~Y.~Liu, J.~Deng, B.~L.~Chen and P.~Yang,
%``Does DaYa-Bay reactor play an important role in theta(13) of lepton mixing
%(PMNS) matrix?,''
arXiv:hep-ph/0409155;
%%CITATION = HEP-PH 0409155;%%
 K.~Hagiwara, in Proceedings of the 
%``Physics prospects of future neutrino oscillation experiments in Asia,''
Fujihara Seminar on Neutrino Mass and Seesaw Mechanism (SEESAW 1979-2004), Ibaraki, Japan, 2004
 [arXiv:hep-ph/0410229]; 
%%CITATION = HEP-PH 0410229;%%
 P.~Huber, M.~Lindner and T.~Schwetz,
%``R2D2 - a symmetric measurement of reactor neutrinos free of systematical
%errors,''
arXiv:hep-ph/0411166;
%%CITATION = HEP-PH 0411166;%%
 C. Albright {\it et al}., 
arXiv:physics/0411123; 
%%CITATION = NONE;%%
J.~E.~Campagne and A.~Cazes,
%``The \theta_{13} and \delta_{CP} sensitivities of the SPL-Frejus project
%revisited,''
arXiv:hep-ex/0411062, and references cited therein..
%%CITATION = HEP-EX 0411062;%%
%
%
%
\bibitem{Maki:1962mu}
Z.~Maki, M.~Nakagawa and S.~Sakata,
%``Remarks On The Unified Model Of Elementary Particles,''
Prog.\ Theor.\ Phys.\  {\bf 28}, 870 (1962);
%%CITATION = PTPKA,28,870;%%
 S.~Eidelman {\it et al.}  [Particle Data Group Collaboration],
%``Review of particle physics,''
Phys.\ Lett.\ B {\bf 592}, 1 (2004).
%%CITATION = PHLTA,B592,1;%%
%
%
%
\bibitem{Joshipura:2004ws}
A.~S.~Joshipura, in Proceedings of the 
Neutrino Oscillations and their Origin (NOON 2004), Tokyo, Japan, 2004
%``Summary of Model Predictions for $U_{e3}$,''
 [arXiv:hep-ph/0411154].
%%CITATION = HEP-PH 0411154;%%
%
%
%
\bibitem{Athar:2002uj}
See, for instance, H.~Athar, in Proceedings of the
6th Constantine High-Energy Physics School on Weak and 
Strong Interactions Phenomenology, Constantine, Algeria, 2002, 
 edited by N. Mebarki and J. Mimouni
%``Some aspects of neutrino astrophysics,''
 [arXiv:hep-ph/0212387].
%%CITATION = HEP-PH 0212387;%%
%
%
%
\bibitem{Gaisser:1997eu}
T.~K.~Gaisser and T.~Stanev,
%``Path length distributions of atmospheric neutrinos,''
 Phys.\ Rev.\  {\bf D57}, 1977 (1998);
%%CITATION = ASTRO-PH 9708146;%%
 {\em ibid.,}
%``Pathlength Distributions Of Atmospheric Neutrinos,''
 Nucl.\ Phys.\ Proc.\ Suppl.\  {\bf 70}, 335 (1999);
%%CITATION = NUPHZ,70,335;%%
 M.~Honda, T.~Kajita, K.~Kasahara and S.~Midorikawa,
 %``Comparison of 3-dimensional and 1-dimensional schemes in the  calculation of
%atmospheric neutrinos,''
 Phys.\ Rev.\  {\bf D64}, 053011 (2001).
%%CITATION = HEP-PH 0103328;%%
%
%
%
\bibitem{Stecker:1978ah}
V. S. Berezinsky and A. Yu. Smirnov, 
Ap.  Space Sci. {\bf 32}, 461 (1975); 
%%CITATION=NONE;%% 
S.~H.~Margolis, D.~N.~Schramm and R.~Silberberg,
%``Ultrahigh-Energy Neutrino Astronomy,''
Astrophys.\ J.\  {\bf 221}, 990 (1978);
%%CITATION = ASJOA,221,990;%%
F.~W.~Stecker,
%``Diffuse Fluxes Of Cosmic High-Energy Neutrinos,''
Astrophys.\ J.\  {\bf 228}, 919 (1979);
%%CITATION = ASJOA,228,919;%%
V.~S.~Berezinsky and V.~V.~Volynsky, in 
{\it Proceedings of the 
  16th International Cosmic Ray Conference, 
 Kyoto, Japan, 1979}, Vol. 10, p. 332;
G.~Domokos, B.~Elliott and S.~Kovesi-Domokos,
%``Cosmic neutrino production in the Milky Way,''
J.\ Phys.\ G {\bf 19}, 899 (1993);
%%CITATION = JPHGB,G19,899;%%
V.~S.~Berezinsky, T.~K.~Gaisser, F.~Halzen and T.~Stanev,
%``Diffuse radiation from cosmic ray interactions in the galaxy,''
Astropart.\ Phys.\  {\bf 1}, 281 (1993);
%%CITATION = APHYE,1,281;%%
 G.~Ingelman and M.~Thunman,
%``Particle Production in the Interstellar Medium,''
arXiv:hep-ph/9604286.
%%CITATION = HEP-PH 9604286;%%
%
%
%
\bibitem{Athar:2000ak}
H.~Athar,
%``Neutrino conversions in cosmological gamma-ray burst fireballs,''
Astropart.\ Phys.\  {\bf 14}, 217 (2000).
%%CITATION = HEP-PH 0004191;%%
%
%
%
\bibitem{Gondolo:1995fq}
M.~Thunman, G.~Ingelman and P.~Gondolo, 
%``Charm production and high energy atmospheric muon and neutrino fluxes,''
Astropart.\ Phys.\  {\bf 5}, 309 (1996).
%%CITATION = HEP-PH 9505417;%%
%
%
%
\bibitem{uno}
 See, for instance, {\tt http://ale.physics.sunysb.edu/uno/}.
 Discussion of some other proposed Megaton 
detectors such as the Hyper-Kamiokande and the Megaton 
project at Fr\'{e}jus is available from web site of 
{\tt UNO Collaboration Meeting: October 14$-$16, 2004, Colorado}.  
%
%
%
\bibitem{book}
 T. K. Gaisser, {\em Cosmic Rays and Particle Physics}
 (Cambridge University Press, Cambridge, UK, 1990), p. 85.
 See, also, T.~Stanev,
%``Possible tau appearance experiment with atmospheric neutrinos,''
Phys.\ Rev.\ Lett.\  {\bf 83}, 5427 (1999).
%%CITATION = ASTRO-PH 9907018;%%
%
%
%
\bibitem{cngs}
See, {\tt http://proj-cngs.web.cern.ch/proj-cngs/}.
%
%
%
\bibitem{amanda}
{\tt http://amanda.physics.wisc.edu/}.
%
%
%
\bibitem{baikal}
{\tt http://www.ifh.de/baikal/baikalhome.html}.
%
%
%
\end{thebibliography}
\end{document}